\documentclass[onecolumn]{aastex631}

\DeclareUnicodeCharacter{02BC}{}

\begin{document}

\title{Identification of Carbon Stars from LAMOST DR7}

\author[0000-0003-1454-2268]{Linlin Li}
\affiliation{Department of Physics, Hebei Normal University, Shijiazhuang 050024, China; \url{wenyuancui@126.com}}
\author{Kecheng Zhang}
\affiliation{Department of Physics, Hebei Normal University, Shijiazhuang 050024, China; \url{wenyuancui@126.com}}

\author[0000-0002-0786-7307]{Wenyuan Cui}
\affiliation{Department of Physics, Hebei Normal University, Shijiazhuang 050024, China; \url{wenyuancui@126.com}}

\author[0000-0002-0349-7839]{Jianrong Shi}
\affiliation{Key Lab of Optical Astronomy, National Astronomical
  Observatories, Chinese Academy of Sciences \\
A20 Datun Road, Chaoyang, Beijing 100101, China}

\author[0000-0002-9468-2502]{Wei Ji}
\affiliation{Key Laboratory of Space Astronomy and Technology, National Astronomical Observatories, CAS, \\
Beijing 100101, China}

\affiliation{School of Astronomy and Space Science, University of Chinese Academy of Sciences, Beijing 100049, Peopleʼs Republic of China}
\author{Zhenyan Huo}
\affiliation{Department of Physics, Hebei Normal University, Shijiazhuang 050024, China; \url{wenyuancui@126.com}}
\author{Yawei Gao}
\affiliation{Department of Physics, Hebei Normal University, Shijiazhuang 050024, China; \url{wenyuancui@126.com}}
\author[0000-0003-2413-9587]{Shuai Zhang}
\affiliation{Department of Physics, Hebei Normal University, Shijiazhuang 050024, China; \url{wenyuancui@126.com}}

\author[0000-0002-2473-9948]{Mingxu Sun}
\affiliation{Department of Physics, Hebei Normal University, Shijiazhuang 050024, China; \url{wenyuancui@126.com}}


\begin{abstract}

Carbon stars are excellent kinematic tracers of galaxies and play important roles in understanding the evolution of the Galaxy. Therefore, it is worthwhile to search for them in a large amount of spectra. In this work, we build a new carbon star catalog based on the LAMOST DR7 spectra. The catalog contains 4542 spectra of 3546 carbon stars, identified through line index and near-infrared color-color diagrams. Through visual inspection of the spectra, we further subclassify them into 925 C--H, 384 C--R, 608 C--N, and 1292 Ba stars. However, 437 stars could not be sub-classified due to their low signal-to-noise. Moreover, by comparing with LAMOST DR7 pipeline we find 567 more carbon stars and visually sub-classify them. We find that on the $J-H$ vs. $H-K_{\rm s}$ two-color diagram, C--N stars can be reliably distinguished from the other three sub-types. Additionally, by utilizing the Gaia distance, we study the distribution of carbon stars in the H-R diagram and identify 258 dwarf carbon stars by the criterion $M_{\rm G}>$5.0\,mag. Finally, we present the spatial distribution in Galactic coordinates of the 3546 carbon stars. The majority of C-N, C-R, and Ba stars are distributed at low Galactic latitudes, while most C--H and dC stars distribute at high Galactic latitudes. 

\end{abstract}
 
\section{Introduction}

Carbon stars are rare and peculiar objects first discovered by \citet{1869AN.....73..129S}, and characterized by strong carbon molecular bands, such as CH, CN, and C$_2$, in their optical spectra. They usually have similar temperature and luminosity with late G, K and M stars, while showing higher carbon abundance than oxygen (i.e.,C/O$>1$) \citep{1993PASP..105..905K}. The atmospheric carbon of dwarf carbon stars is thought to be obtained by mass transfer from an asymptotic giant branch (AGB) companion, which is now a white dwarf star \citep{1983ARA&A..21..271I,1988ApJ...328..653B}. Cool and luminous carbon stars acquire their carbon enrichment through the ongoing third dredge-up during the AGB phase.

According to their different spectral features, carbon stars can be classified into different sub-classes \citep{1993PASP..105..905K,1996ApJS..105..419B}, including C--H, C--R, C--N, C--J and Ba stars. The classification criteria have been discussed by \cite{2018ApJS..234...31L}. They can also be divided into giant and dwarf carbon stars (dC) depending on their brightness. Note that these classifications have no direct relationship with their origin. It is well known that most stars belong to binary systems \citep{2013ARA&A..51..269D}and they are no exception for carbon stars. Almost all Ba and C--H stars are binaries \citep{1980ApJ...238L..35M,1983ApJ...268..264M,2016A&A...586A.158J,2023A&A...671A..97E,2023AJ....165..154G}. \cite{1984JRASC..78S.205A} carried out a series of radial velocity monitoring for C--N stars, and found signs of binary companions for C--N stars. However, \citet{1997PASP..109..256M} did not find any evidence of binary for 22 C--R stars after 16 years of radial velocity monitoring.

Carbon dwarfs have been proven to be the most common type \citep{2000IAUS..177...27G}, but the discovery of carbon dwarfs occurred much later than that of carbon giants. Most of the known carbon stars so far are giants due to the lower luminosity of dC stars. \citet{1977ApJ...216..757D} discovered a carbon star (G77-61), which, based on its parallax and spatial motion, is a low-luminosity object located on the M-dwarf sequence in the H-R diagram, and G77-61 now appears to be a prototype of carbon dwarfs.  More dC stars are needed to explain their origins and the relationship with the carbon giants. 

Among the sub-types of carbon stars, the C–N and C–H stars receive more attention. Since C–N stars are mostly in the AGB stage, they are intrinsically bright and can be easily identified by the strong carbon molecular bands in their spectra. They have been used as kinematic probes of the Galaxy at large distances \citep{2007A&A...473..143D,2013Ap.....56...68B}. Additionally, the small dispersion in the absolute magnitude of C–N stars allows them to be used as reliable candles to measure the distance of extragalaxies \citep{2020MNRAS.495.2858R,2023MNRAS.522..195P}. C–H stars are metal-poor and can be found mostly in halo populations \citep{2005MNRAS.359..531G}. A larger sample of C–H stars is very helpful in improving our knowledge about the formation history of the stellar halo. Moreover, their more metal-poor counterparts, carbon-enhanced metal-poor (CEMP) stars \citep{2016A&A...587A..50A,2022ApJ...934..110S}, provide essential information about the early Galactic chemical evolution and the nature of the very first stars.

So far, many works have been carried out for systematically identifying carbon stars, e.g., \citet{2001BaltA..10....1A} published  a catalog of 6891 carbon stars, which is an updated and revised version of Stephenson's catalog of Galactic cool carbon stars \citep{1973gcss.book.....S,1989PW&SO...3...53S}, \citet{1998MNRAS.294....1T} reported 48 cool (N-type) carbon stars, and \citet{2001A&A...375..366C} presented a sample of 403 faint high-latitude carbon (FHLC) stars selected from the digitized objective plasma plates of the Hamburg/ESO Survey (HES). Over the next ten years, several studies systematically searched for carbon stars from the Sloan Digital Sky Survey (SDSS) \citep{2002AJ....124.1651M,2004AJ....127.2838D,2013ApJ...765...12G,2014SCPMA..57..176S}. Restricted to the sky survey area of SDSS, the identified carbon stars are mainly FHLC stars. Since the data release of the Large Sky Area Multi-Object Fiber Spectroscopy Telescope (LAMOST) in the year 2011, new carbon stars were reported from its massive number of spectra, e.g., \citet{2015RAA....15.1671S} used machine learning to identify 183 carbon stars from the LAMOST pilot survey data, \citet{2016ApJS..226....1J} obtained a sample of 894 carbon stars from LAMOST DR2 using multiple line index spaces and $JHK_s$ color-magnitude diagram, and \citet{2018ApJS..234...31L} reported  2651 carbon stars from LAMOST DR4 using machine learning.

In this paper, our goal is to use the LAMOST DR7 dataset to expand the sample of carbon stars. This paper is organized as follows. In Section~ 
\ref{sec:LAMOST}, we introduce the dataset of LAMOST DR7. In Section~\ref{sec:identification steps of carbon stars}, we describe the identification method and the process for searching for carbon stars, and we provide the final results. In section~\ref{Discussion}, we discuss the distribution of carbon stars in the {\it J}$-${\it H} vs. {\it H}$-${\it K}$_{\mathrm s}$ and {\it H}$-${\it R} diagrams, in which we identify 258 possible dC stars, and the spatial distribution of the carbon stars is also shown in this section. Finally, a brief summary is given in Section~\ref{sec:conclusion}.

\section{Data} \label{sec:LAMOST}
\subsection{LAMOST DR7}

LAMOST, also called the Guo Shou Jing telescope, is a 4 m reflective Schmidt telescope with 4000 fibers in a 20 square degree focal plane \citep{2012RAA....12.1197C,2012RAA....12..723Z}. LAMOST can simultaneously  obtain the spectra of 4000 targets in a single exposure, and its resolution is $\sim$1800 for the low resolution mode.

In June 2019, LAMOST completed its seventh year of sky survey. On March 31, 2020, the LAMOST DR7 dataset, including the pilot survey and the seven years regular survey, was officially released to astronomers. In this release, it comprises 4922 low resolution observation plates. The DR7 dataset contains 14,487,406 spectra, including 10,599,979 low resolution spectra, and 1,008,710 nontime domain and 2,878,717 time domain medium resolution spectra. In the LAMOST DR7 Low-Resolution Spectroscopic (LRS) General Catalog\footnote{http://dr7.lamost.org/catalogue}, there are 4402 spectra of 3565 stars labeled as carbon for the ``subclass" parameter (stellar spectral type). However, the contamination rate is not presented, and no classification of the carbon stars is given. In this work, we systematically search for different types of carbon stars from more than 10 million low resolution spectra.

\subsection{Line indices}

We use the multiple line index spaces to select carbon stars, as the spectra of carbon stars are characterized by strong absorption bands of carbon-containing molecules, such as CH, C$_2$ and CN.

The line index is defined by the following equation according to the equivalent width \citep{1994ApJS...94..687W,2015RAA....15.1137L}:
\begin{equation}
EW=\int[1-\frac{f_{line}(\lambda)}{f_{cont}(\lambda)}]d\lambda,
\end{equation}
where, $f_{cont}(\lambda)$ and $f_{line}(\lambda)$ are the fluxes of the continuum and the spectral line, respectively, both of them are functions of wavelength. The line index under this definition is in \AA. We measure five line indices including C$_2$($\lambda$\,4737\,\AA, $\lambda$\,5165\,\AA, and $\lambda$\,5635\,\AA) and CN ($\lambda$\,7065\,\AA, and $\lambda$\,7820\,\AA). The definition of line indices is listed in Table~\ref{tab:table1}.
\begin{deluxetable}{lcl}
\tablenum{1}
\tablecaption{Line Indices Definition in This Work\label{tab:table1}}
\tablewidth{0pt}
\tablehead{
\colhead{Name} & \colhead{Index Bandpass (\AA)} & \colhead{Pseudo-continua (\AA)} }
\decimalcolnumbers
\startdata
C$_2$(4737\AA) & 4620--4742 & 4580--4620\quad 4742--4812  \\
C$_2$(5165\AA) & 4980--5170 & 4930--4980\quad 5170--5235  \\
C$_2$(5635\AA) & 5350--5640 & 5300--5350\quad 5640--5700 \\
CN(7065\AA) & 7065--7190 & 7025--7065\quad 7190--7230 \\
CN(7820\AA) & 7820--8000 & 7790--7820\quad 8000--8040 \\
\enddata
\end{deluxetable}

\section{Identification and Spectral classification of carbon stars} \label{sec:identification steps of carbon stars}
\subsection{The identification steps of carbon stars}\label{sec:the idendification of carbon stars}
The LAMOST DR7 dataset contains 10,599,979 low resolution spectra. In order to derive the carbon star candidates, we use multiple line index spaces to remove as many contaminants as possible. 

\begin{figure}[!ht]
\plotone{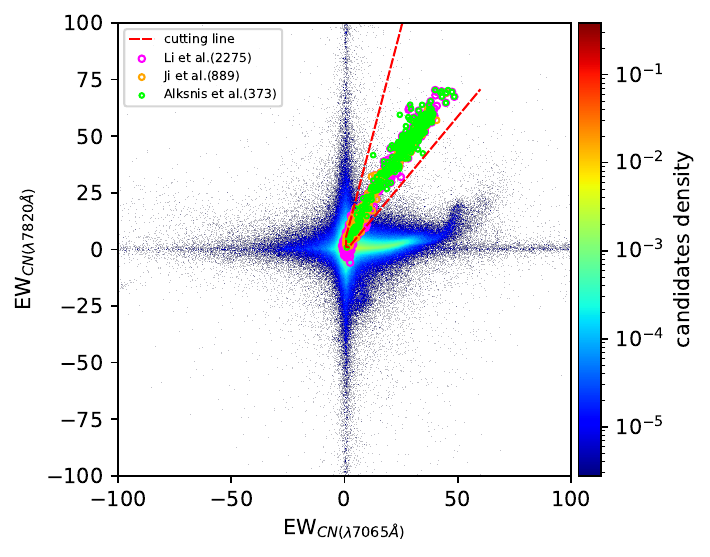}
\caption{Distribution of the 9,060,920 spectra and the known carbon stars in EW $_{{\rm CN}(\lambda\,7820\,{\rm \AA})}$ vs. EW$_{{\rm CN} (\lambda\,7065\,{\rm \AA})}$ plane. The solid dots indicate the density of 9,060,920 candidates with S/N(i)$>$10. The red dashed lines indicate the cuts used to remove contaminants, and the magenta, orange, and green unfilled circles represent the carbon star samples from \citet{2018ApJS..234...31L}, \citet{2016ApJS..226....1J}, and \citet{2001BaltA..10....1A}, respectively. The numbers in brackets indicate the number of known carbon stars in each sample.  
\label{fig:figure1.0}}
\end{figure}

The high signal-to-noise spectra are helpful to us to identify and classify carbon stars more accurately. Therefore, we only selected the spectra of S/N (i)$>$10 from LAMOST DR7, and obtained 9,060,920 stellar spectra, in which some objects have been observed in more than one epoch. In order to determine the intrinsic location of carbon stars in line index spaces, we cross-matched the known carbon star catalogs of \citet{2018ApJS..234...31L}, \citet{2016ApJS..226....1J}, and \citet{2001BaltA..10....1A} with the selected  LAMOST DR7 dataset, and found 2275, 889, and 373 carbon stars, respectively. In Figure~ \ref{fig:figure1.0}, we draw all the known carbon stars and candidates in the EW$_{{\rm CN}(\lambda\,7820\,{\rm \AA})}$ vs. EW$_{{\rm CN}(\lambda\,7065\,{\rm \AA})}$ plane, and it can be seen that the distribution of known carbon stars spreads from the center to the upper right in a narrow area. In the central region around the origin point in Figure~\ref{fig:figure1.0}, different types of stars are mixed, and it is very difficult to identify the carbon stars in the highly contaminated region. Hence, we exclude the central region in the following discussion. We adopt the following empirical criteria in the EW$_{{\rm CN}(\lambda\,7820\,{\rm \AA})}$ vs. EW$_{{\rm CN}(\lambda\,7065\,{\rm \AA})}$ plane so that most of the known carbon stars are included. 
After this step, we get 722,171 candidates, accounting for about 8\% of the total number before cutting. The criteria are: 

\begin{equation}
\begin{array}{l}
1.2 \times EW_{{\rm CN}(\lambda\,7065\,{\rm \AA})} -1.29 < EW_{{\rm CN}(\lambda\,7820\,{\rm \AA})}\\ 
< 3.8 \times EW_{{\rm CN}(\lambda\,7065\,{\rm \AA})} + 3.0,\\
EW_{{\rm CN}(\lambda\,7065\,{\rm \AA})} > 0.7, \\
EW_{{\rm CN}(\lambda\,7820\,{\rm \AA})}>1.0. 
\end{array}
\label{eq2}
\end{equation}


\begin{figure}[!ht]
\plotone{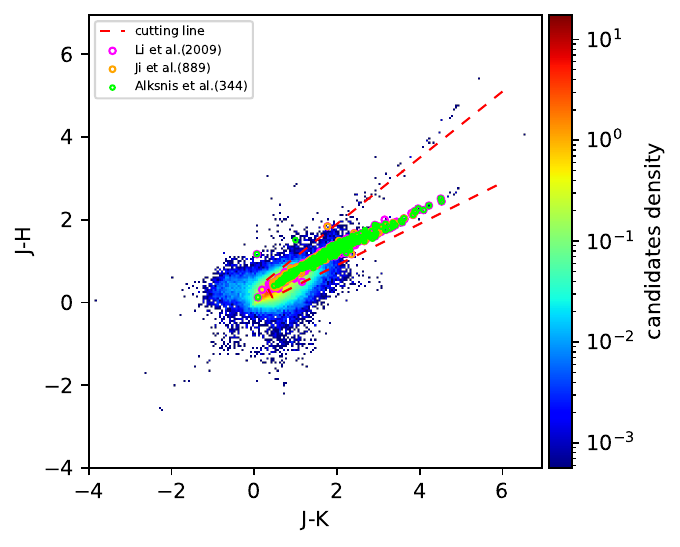}
\caption{Distribution of the 706,597 candidates and the known carbon stars in the {\it J-H} vs. {\it J-K$_{\rm s}$} color-color diagram. The solid dots indicate the density of the candidates. The other symbols are the same as in 
Figure~\ref{fig:figure1.0}.  
\label{fig:figure3}}
\end{figure}

We further remove pollution based on the infrared color-color diagram, since previous works have shown that the distribution of carbon stars in the {\it J$-$H} vs. {\it H$-$K$_{\rm s}$} two-color diagram is highly concentrated \citep{2000MNRAS.314..630T,2012Ap.....55..424G,2015RAA....15.1671S}. The values of {\it J}, {\it H}, and {\it K$_{\rm s}$} for both known carbon stars and candidates are obtained by cross-identifying with the 2MASS All-Sky point source catalog \citep{2006AJ....131.1163S}. However, there are 15,575 candidates with no matching information in the 2MASS database. We visually examined their spectra and identified 68 carbon stars. The {\it J}, {\it H}, and {\it J$-$K$_{\rm s}$} photometric values are available for 706,597 carbon star candidates, which are plotted in the {\it J$-H$} vs. {\it J$-$K$_{\rm s}$} plane in Figure~\ref{fig:figure3}, along with the known carbon star samples. The known carbon stars are primarily concentrated in a narrow branch extending from the center to the upper right (see Figure~\ref{fig:figure3}). To remove pollution, we apply several empirical lines to simultaneously cut the candidates and known carbon star samples. The following empirical criteria are adopted.

\begin{equation}
\begin{array}{l}
J-H>0.5\times(J-K_{\rm s})-0.1, \\
J-H<0.8\times(J-K_{\rm s})+0.3,\\    
J-H>-2.0\times(J-K_{\rm s})+1.0.          
\end{array}
\label{eq3}
\end{equation}

After applying these steps, the number of candidate samples is reduced to 609,702. Although a few known carbon stars fall outside the cutting range, the number of carbon star samples from \citet{2001BaltA..10....1A}, \citet{2016ApJS..226....1J}, and \citet{2018ApJS..234...31L} is 338, 885, and 1895, respectively. Therefore, the impact of these outliers is not significant.

\begin{figure}[!ht]
\plotone{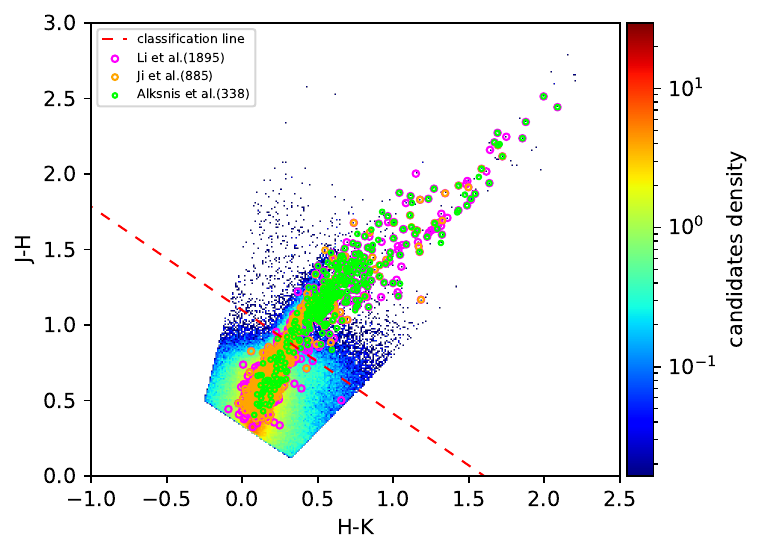}
\caption{Distribution of the 609,702 candidates and the known carbon stars in the $J-
H$ vs. $H-K_{\rm s}$ color-color diagram. The solid dots with marked density 
represent the candidates. The red dashed line (i.e., Equation(\ref{eqjhk}) divides the candidates and known carbon stars into two groups, the Warm Group and the Cool Group. The lower left of the dashed line is the Warm Group, while the upper right is the Cool Group. The other 
symbols are the same as in Figure~\ref{fig:figure1.0}.  
\label{fig:Figure 4}}
\end{figure}

The strong absorption bands, such as CH ($\lambda4300$\,\AA) and C$_2$ ($\lambda$4737\,\AA; $\lambda$5165\,\AA), in the blue bands, are the most obvious features of warm carbon stars (C--R, C--H, and Ba stars). On the other hand, cool carbon stars (C--N stars) show almost no flux in the blue bands. Therefore, it is important to first distinguish between warm and cool carbon candidates before using this characteristic to select carbon stars. In Figure \ref{fig:Figure 4} we map the candidates and known carbon stars in the {\it J}$-${\it H} vs. {\it H}$-${\it K} plane. The red dashed line divides them into two parts, the Warm Group on the lower left, while the Cool Group on the upper right. The equation of the red dashed line is from \citet{2015RAA....15.1671S} as following
 
\begin{equation}
J–H=-0.6851\times(H-K_{\rm s})+1.0974.
\label{eqjhk}
\end{equation}

\subsubsection{The Warm Group}
\label{subsubsec:the warm group}

\begin{figure}[!ht]
\plotone{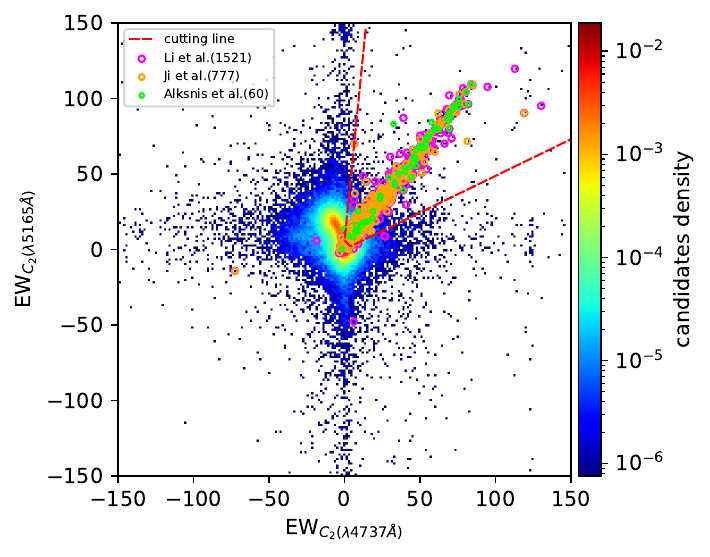}
\caption{Distribution of the 586,073 warm candidates and the known carbon stars in the EW
$_{C_2(\lambda5635\AA)}$ vs. EW$_{C_2(\lambda4737\AA)}$ . The solid 
dots indicate the density of the warm candidates, the red 
dashed lines represent the cutting lines (i.e., Equation (\ref{equ5})), and other 
symbols are the same as in Figure \ref{fig:figure1.0}. 
\label{fig:Figure 5}}
\end{figure}

We present the Warm Group carbon stars in the EW$_{{\rm C}_2(\lambda5635\,{\rm \AA})}$ vs. EW$_{{\rm C}_2(\lambda4737\AA)}$ plane, as shown in 
Figure~\ref{fig:Figure 5}. We can see that the locus of the known carbon stars
extends from the center of the figure to the upper right. Candidates with weak C$_2$($\lambda4737$\,\AA; $\lambda$5165\,\AA) characteristics distributed in the center region are also the major contaminants. In order to remove more contaminants, we abandond the center region. The criteria are:

\begin{equation}
\begin{array}{l}
EW_{{\rm C}_2(\lambda5165\,{\rm \AA})}<10.0 \times EW_{{\rm C}_2(\lambda4737\,{\rm \AA})} +6.0,\\
EW_{{\rm C}_2(\lambda5165\,{\rm \AA})}>0.3 \times EW_{{\rm C}_2(\lambda4737\,{\rm \AA})} – 2.0,\\
EW_{{\rm C}_2(\lambda5165{\rm \AA})}>-1.0 \times EW_{{\rm C}_2(\lambda4737{\rm \AA})} + 6.0.
\end{array}
\label{equ5}
\end{equation}



\begin{figure}[!ht]
\plotone{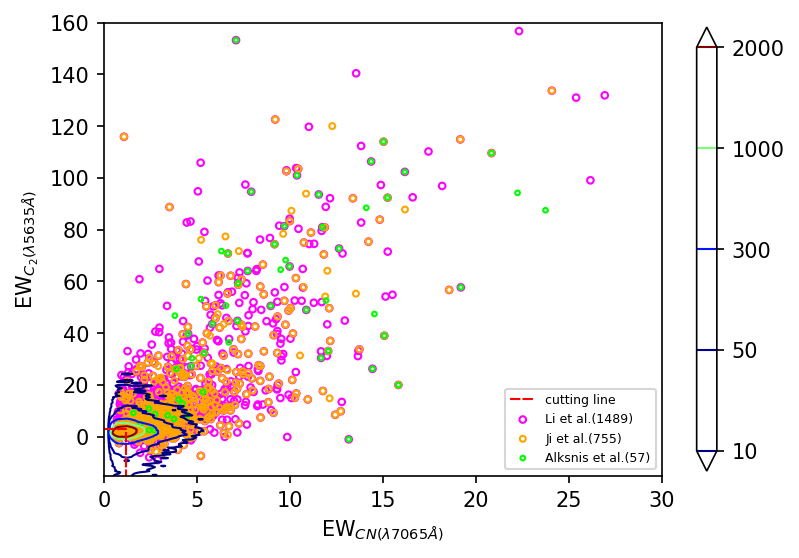}
\caption{Distribution of the 127,943 warm candidates and the known carbon stars in the EW
$_{{\rm C}_2(\lambda5635\,{\rm \AA})}$ vs. EW$_{{\rm CN}(\lambda7065\,{\rm \AA})}$ plane. The contours 
represent the density of the candidates. The red dashed lines are the cutting 
lines (i.e., Equation (\ref{equ6})). Other symbols are the same as in Figure \ref{fig:figure1.0}.  
\label{fig:figure6}}
\end{figure}

Next, we remove as many warm star contaminators as possible in the $EW_{{\rm C}_2(\lambda5635\,{\rm \AA})}$ vs. $EW_{{\rm CN}(\lambda7065\,{\rm \AA})}$ plane. As can be seen in Figure \ref{fig:figure6}, most of the known carbon stars are distributed in the upper right. In order to ensure that we can identify carbon stars visually, we use the empirical Equation (\ref{equ6}) to remove contaminators. After this cut, 85,405 (about 67\%) candidates are left. Finally, we visually identify 3631 spectra of 2903 carbon stars from the remaining 85,405 candidates.

\begin{equation}
\begin{array}{l}
EW_{{\rm CN}(\lambda7065\,{\rm \AA})}<1.2,\\
EW_{{\rm C}_2(\lambda5635\,{\rm \AA})}<3.0.
\end{array}
\label{equ6}
\end{equation}

\subsubsection{The Cool Group} 
\label{subsubsec:the cool group}
\begin{figure}[!ht]
\plotone{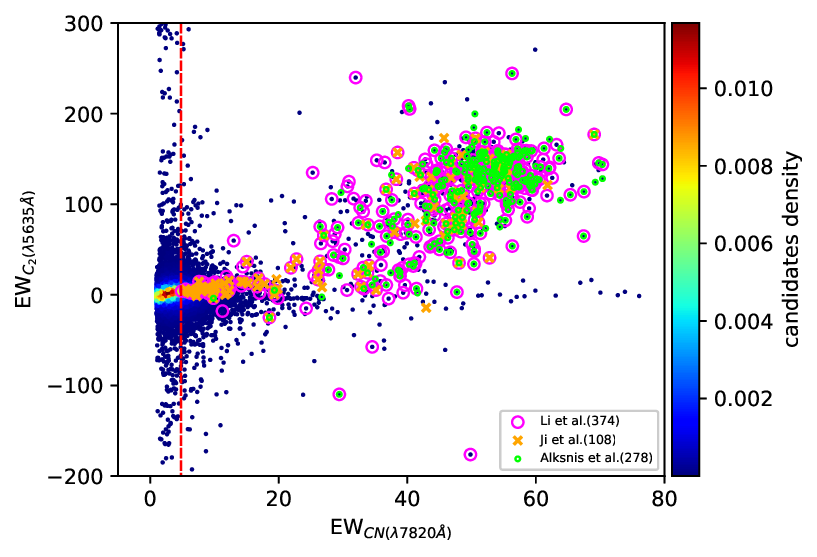}
\caption{The distribution of 23,629 cool candidates and known carbon stars in the EW
$_{{\rm C}_2 (\lambda5635\,{\rm \AA})}$ vs. EW$_{{\rm CN}(\lambda7820\,{\rm \AA})}$ plane. The red dashed line represents the cutting line (i.e., Equation (\ref{equ8})). The dots indicate the density of the candidates. The other symbols 
are the same as in Figure \ref{fig:figure1.0}.
\label{fig:figure7.0}}
\end{figure}

The Cool Group carbon stars generally have lower temperatures, and almost all carbon stars in this group are C--N stars. Although these cool carbon stars have almost no flux at the blue end (at $\lambda<$4400\,\AA, even at $\lambda<$5000\,\AA), they show strong CN absorption in the red bands. As a result, we use the C$_2(\lambda$5635\,\AA) and CN($\lambda7820$\,\AA) molecule lines at the red end to remove contaminators. We display this group, which includes the 23,629 candidates and the known carbon star samples, in the $EW_{{\rm C}_2(\lambda5635\,{\rm \AA})}$ vs $EW_{{\rm CN}(\lambda7820\,{\rm \AA})}$ plane. As shown in Figure \ref{fig:figure7.0}, all the known carbon stars are located in the region where $EW_{{\rm CN}(\lambda7820\,{\rm \AA})}$ is larger than a certain value, so we use Equation (\ref{equ8}) to remove contaminants. 

\begin{equation}
EW_{{\rm CN}(\lambda7820\,{\rm \AA})}>4.8.
\label{equ8}
\end{equation}

After this cut, 18,325 (about 77.6\%) contaminants are removed. Finally, 769 spectra of 575 carbon stars are identified by eye inspection from the remaining 5304 candidates.

\subsection{Spectral classification} \label{subsec:spectral classification}

For the spectral classification of carbon stars in this work, we adopt the latest carbon star classification system 
published by \citet{1996ApJS..105..419B} which divides carbon stars into mainly five types: C--R, C--H, C--N, C--J, and Ba 
stars. Since the isotope ratio $^{12}$C/$^{13}$C cannot be derived from the low-resolution spectrum, we cannot identify the C--J stars.

To classify the spectra of carbon stars into different sub-types, we need to understand the spectral characteristics of each sub-type, which are summarized by
\citet{2015RAA....15.1671S}, \citet{2016ApJS..226....1J}, and \citet{2018ApJS..234...31L}:
\begin{enumerate}
\item C--H stars: There is a strong G-band at $\lambda4300$\,\AA. The 
secondary P-branch head near $\lambda4342$\,\AA\ is the most significant feature, and the \ion{Fe}{1} line at $\lambda4383$\,\AA\ is weak, they can be used to distinguish C--H stars from C--R and Ba stars. The \ion{Ba}{2} lines at $\lambda4554$\,\AA\ 
and $\lambda6496$\,\AA\ are very strong.
\item C--R stars: Like the strong absorption CN band at $\lambda4215$\,\AA, the strong 
\ion{Ca}{1} line at $\lambda4226$\,\AA\ is also a significant feature of C--R stars. The \ion{Ba}{2} lines at $\lambda4554$\,\AA\ and $\lambda6496$\,\AA\ are very weak. 
\item C--N stars: There is almost no flux at $\lambda<$4400\,\AA\ or even $
\lambda<$5000\,\AA\, which is an important feature to distinguish C--N stars from C--H and C--R stars. The isotopic bands are consistently weak, while lines of s-process elements, particularly Ba, are more enhanced than those in C--R stars.
\item Ba stars: Because Ba stars have strong enhancement of the s-process elements, the \ion{Ba}{2} line at $\lambda$4554\,\AA\ and \ion{Sr}{2} line at $\lambda4077$\,\AA\ are very strong.
\end{enumerate}

\begin{figure*}[!ht]
\plotone{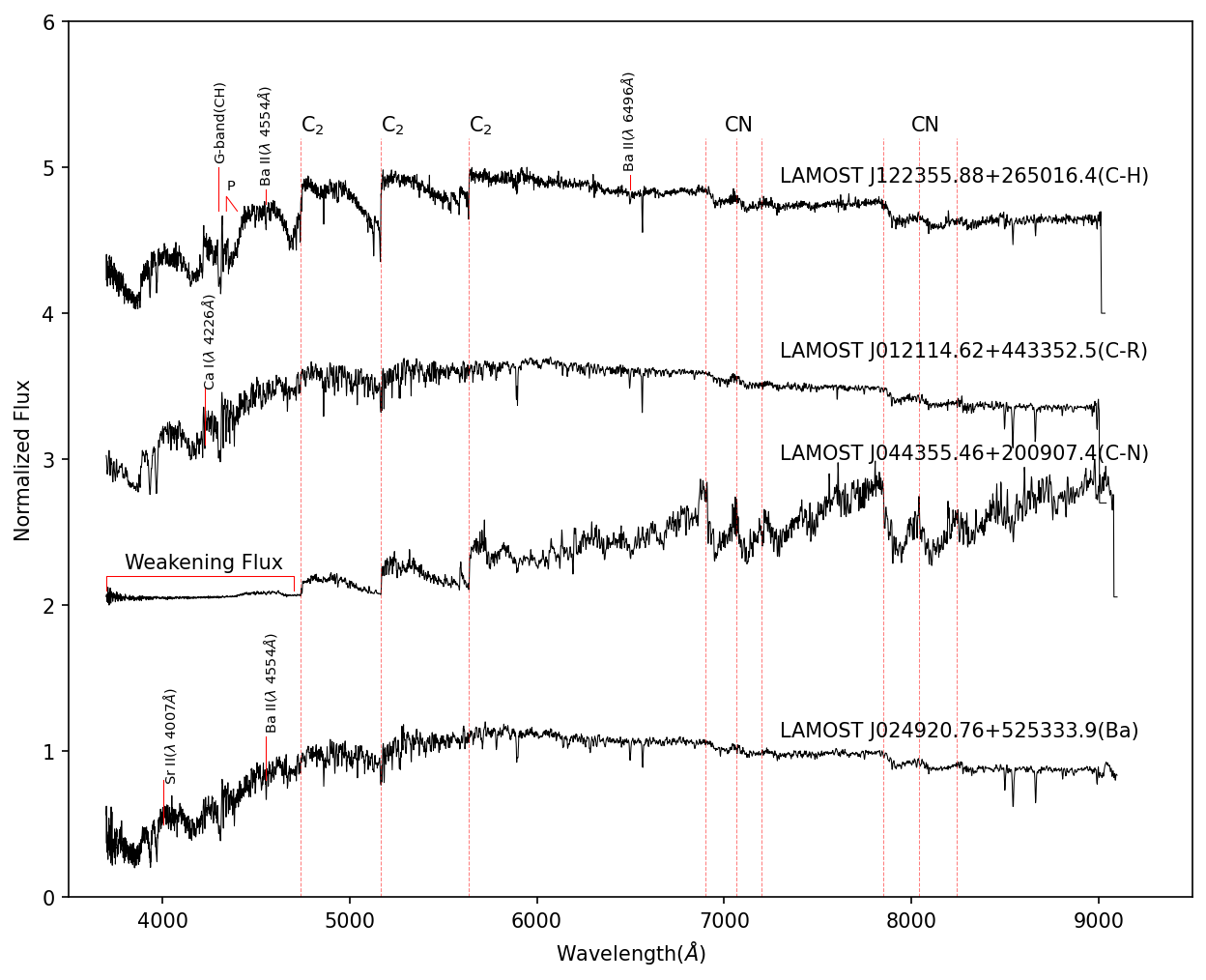}
\caption{The normalized spectra of four LAMOST carbon stars come from this 
work, including a C--H star, a C--R star, a C--N star and a Ba star. 
The G-branch, P-branch band, and Ba II lines are marked for the C--H star by red color. For the spectrum of a C-R star, we label the Ca I line. The weakening flux in the C--N spectrum is labeled in the blue part. The Ba II and Sr II lines are also marked by red color for the Ba star. The red dashed lines represent C2 bands in the blue part and CN bands in the red part. \label{fig:figure7}}
\end{figure*}

\begin{figure*}[!ht]
\plotone{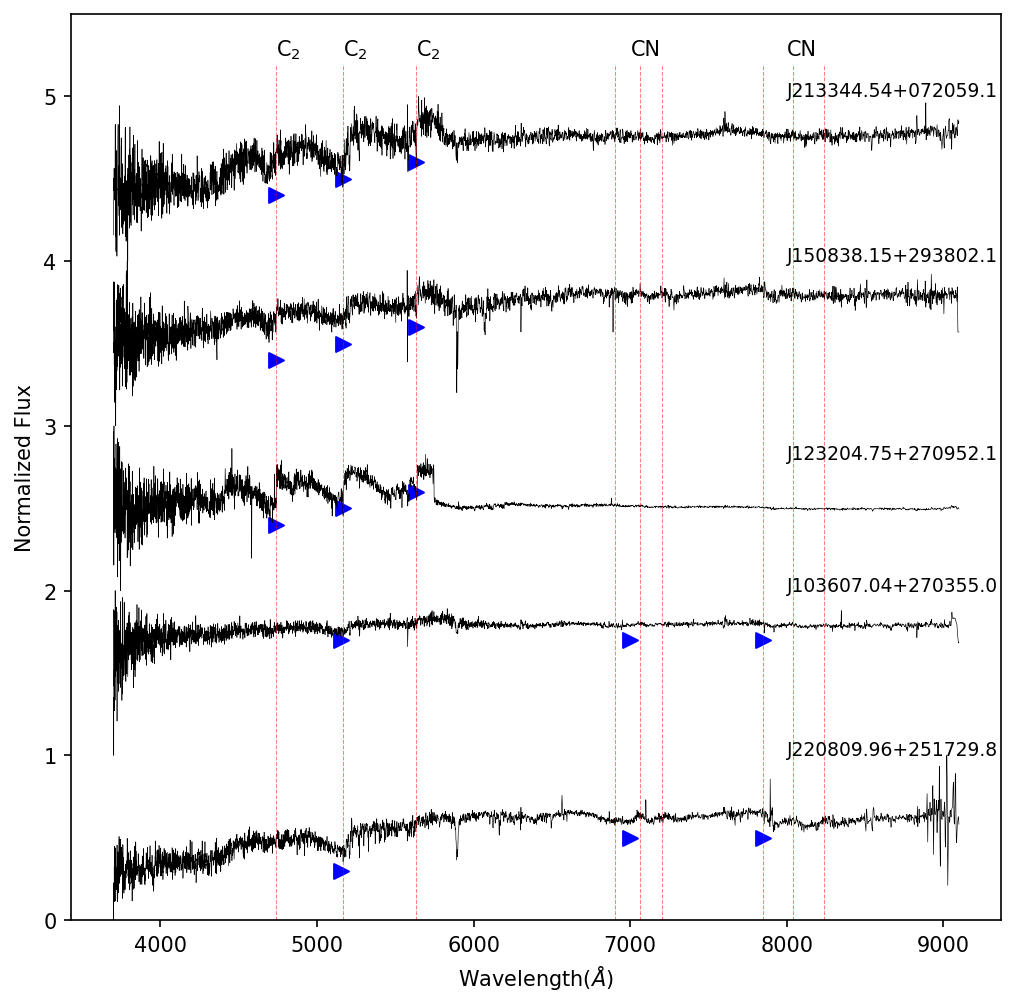}
\caption{Normalized LAMOST spectra of five carbon stars classified as ``UNKNOWN''. The red dashed lines represent the C2 bands in the blue part and CN bands in the red part. The blue triangles represent the carbon features we are using to call these carbon stars.}
\label{fig:figure15}
\end{figure*}

According to these characteristics, we manually classify 4542 spectra of 3546 carbon stars identified in LAMOST DR7. The classification results are shown in Table \ref{tab:table2}. Figure~\ref{fig:figure7} shows four typical LAMOST spectra of carbon stars 
classified by us, which are C--H, C--R, C--N and Ba stars, respectively. Some important characteristic lines and carbon molecular bands are marked. Among the 3546 carbon stars, 437 are defined as ``UNKNOWN'', which means that we cannot classify them by spectra, Figure~\ref{fig:figure15} presents five LAMOST spectra of them, and it is obvious that they are carbon stars. However, their molecular 
band characteristics are fuzzy, and we cannot classify them.

\subsection{Identification results} \label{subsubsec:identification results}

\begin{figure}[!ht]
\plotone{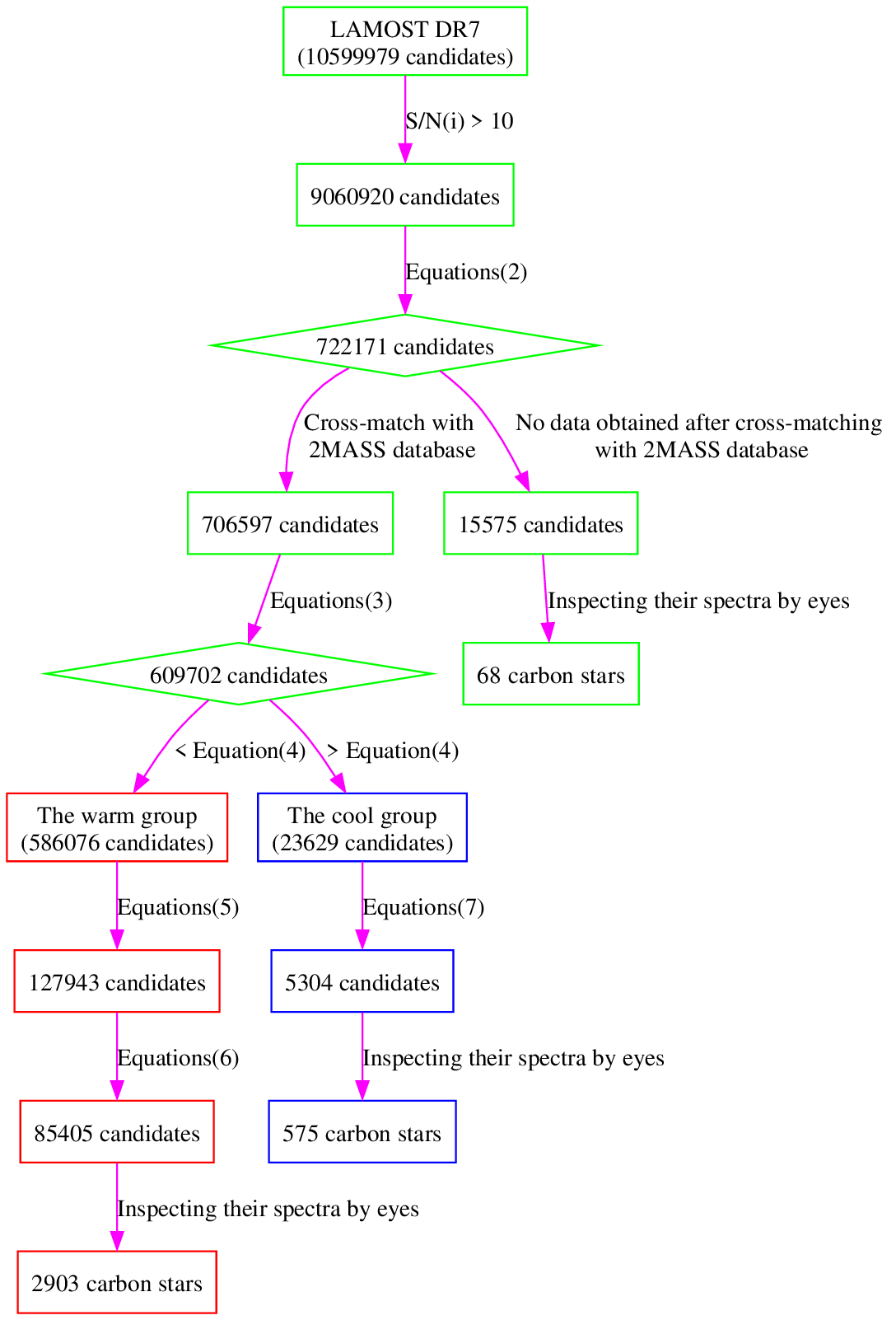}
\caption{Steps of carbon stars identification.}. 
\label{fig:Figure 8}
\end{figure}

By searching for the warm group and the cool group, we identified 2903 carbon stars belonging to the hot group, while 575 carbon stars belonged to the cool group. Additionally, we identified 68 carbon stars in the candidates with no matching information in the 2MASS database. The specific identification steps are summarized in Figure~\ref{fig:Figure 8}, and the final result is presented in Table~\ref{tab:table2}. The information of some example carbon stars is shown in Table~\ref{tab:table3}.

\begin{deluxetable}{lcc}
\tablenum{2}
\tablecaption{Carbon Stars in This Work\label{tab:table2}}
\tablewidth{0pt}
\tablehead{
\colhead{Subtype} & \colhead{Number of Carbon} & \colhead{Number of Carbon}\\
\colhead{} & \colhead{Stars} & \colhead{Stars Spectra} }
\startdata
C--H & 925 & 1217  \\
C--R & 284 & 341  \\
C--N & 608 & 799 \\
Ba & 1292 & 1657 \\
UNKNOWN & 437 & 528 \\
\cline{1-3}
Total Number & 3546 &4542\\
\enddata
\end{deluxetable}

\begin{table*}[!ht]
\centering
 \tablenum{3}
 \caption{Carbon Stars Discovered in LAMOST DR7 
 \label{tab:table3}}

  \resizebox{\textwidth}{!}{
    \begin{tabular}{lllllllllll}
    \hline \hline
        obsid & designation & ra (degree) & dec (degree) & subclass\tablenotemark{a} & $M_{\rm G}$ (mag) & $(G_{\rm BP}$-$G_{\rm RP})_{0}$ (mag) & sp\_type\tablenotemark{b} & dwarf\tablenotemark{d}  \\ \hline
        492105114 & J000021.94+251921.8 & 0.09115 & 25.322726 & Carbon & 0.106571964 & 1.111763218 & C-H & 0 \\ 
        182607108 & J000043.94+463317.7 & 0.183084 & 46.554929 & Carbon & 1.387818663 & 1.16221294 & Ba & 0 \\ 
        269401237 & J000043.99+544642.3 & 0.183294 & 54.778429 & Carbon & -0.541169969 & 1.29267988 & Ba & 0 \\ 
        248409244 & J012137.11+513725.5 & 20.404644 & 51.623759 & Carbon & -2.167650566 & 2.586834633 & C-N & 0 \\ 
        368508222 & J012150.43+011301.3 & 20.4601405 & 1.217029 & Carbon & 9.61557705 & 1.783483731 & C-N & 1 \\ 
        407815063 & J012153.97+411341.2 & 20.474905 & 41.228134 & Carbon & 1.209880871 & 1.260210652 & Ba & 0 \\ 
        470104224 & J012300.73+190749.0 & 20.75305 & 19.1302973 & Carbon & 4.082624671 & 1.035243017 & C-R & 0 \\ 
        393004088 & J012302.65+465155.3 & 20.761044 & 46.865362 & G5 & -0.707715317 & 1.48405494 & Ba & 0 \\
         69216250 & J004637.24+401824.8 & 11.655199 & 40.306906 & Carbon & 8.980534815 & 1.803201731 & UNKNOWN\tablenotemark{c} & 1 \\ \hline
    \end{tabular}}
    \tablenotetext{a}{Spectral sub-type given by LAMOST DR7 pipline}
    \tablenotetext{b}{Spectral sub-type.}
    \tablenotetext{c}{Carbon stars with no identifiable sub-type.}
    \tablenotetext{d}{1 means that the carbon star is a dwarf, while 0 for a giant.}

\end{table*}

\section{Discussion}
\label{Discussion}
\subsection{Compare with the identification results of LAMOST DR7 pipline and LAMOST DR4 catalog}


In the LAMOST DR7 LRS General Catalog, there are 3565 stars (4420 spectra) have been marked as ``carbon''. Firstly, 84.24\% of our carbon stars have been correctly classified as carbon
stars by the LAMOST DR7 pipeline, and the other 15.76\% (559)are only recognized by our method. Among these incorrectly classified carbon stars, 95.53\% (534) have been classified as G5--type stars, and others have been categorized as K5--type star (one), K4--type star (six), K1--type stars (one), and unknown (17), respectively. In our carbon catalog, these stars are classified as 520 Ba, 13 C--N, 12 C--H, 10 C--R, and 4 UNKNOWN, respectively. 

Then we checked the stars (578) that were marked as "carbon" by the LAMOST DR7 pipeline but not by us, using our identification steps in Section \ref{sec:identification steps of carbon stars}. Most of them (81\%) were removed by Equation (\ref{eq2}), since they have relatively weak ${{\rm CN}(7280\,{\rm \AA})}$ and ${{\rm CN}(7065\,{\rm \AA})}$ line. The remaining 19\%  have {\it J}, {\it H}, and {\it J$-$K$_{\rm s}$} photometric, while they all have small line index, and generally fall close to (0,0) in the line index planes, therefore they are discarded when cutting carbon star candidates. We visually inspected these stars, 567 of them are confirmed as true carbon stars, as shown in Table \ref{tab:table4}. About half of them (279) are classified as C--H stars, while, there are 77, 46, 33, and 132 stars categorized as Ba, C--N, C--R stars, and "UNKNOWN", respectively. The S/Ns are very low for the remaining 11 stars, so it's not sure if they are carbon stars.

\begin{table*}
 \tablenum{4}
 \centering
 \caption{Sub-type of Carbon Stars Marked as "carbon" by LAMOST DR7 Pipeline 
  \label{tab:table4}}

\begin{tabular}{lllll}
 \hline
 \hline
  obsid & designation & ra (degree) & dec (degree) & sp\_type  \\ 
  \hline
   616307219 & J031910.17+620020.8 & 49.792375 & 62.005787 & C-N\\
  645002216 & J124804.83+594334.1 & 192.0201253 & 59.7261532 & C-H\\
  199404230 & J060920.95+593107.1 & 92.337312 & 59.518646 & Ba\\
  617401149 & J011427.00+590944.1 & 18.612503 & 59.162267 & C-H\\
  157909037 & J040030.35+584828.8 & 60.126463 & 58.808008 & Ba\\
  583213247 & J133536.72+574224.0 & 203.900881 & 57.706683 & C-H\\
  32208015 & J115227.75+571342.9 & 178.115656 & 57.228607 & UNKNOWN\\
 \hline
\end{tabular}

\end{table*}

Comparing with LAMOST DR4 carbon star catalog of \citep{2018ApJS..234...31L}, we find that there are 764 stars included in the LAMOST DR4 catalog, while they are not contained in  \citep{2018ApJS..234...31L}. There are 537 stars included in \cite{2018ApJS..234...31L} , while we have not found. The reason is that those stars have weak CN bands, therefore, we have omitted them as discussed in Section \ref{sec:identification steps of carbon stars}. Among these stars, 249 objects have also been also identified as carbon stars by the LAMOST DR7 pipeline.

Because we can not know definitively the number of carbon stars in the massive LAMOST DR7 data set, it is impossible to accurately estimate the completeness and contamination rates of our method. However, we can roughly evaluate the upper limit of completeness as $3546/(3546+567) \approx 86.21\%$. Since the carbon stars in our work were visually confirmed, the contamination rate is very low.

\subsection{Distribution of our carbon stars in the {\it J}$-${\it H} vs. $H-K_{\rm s}$ diagram}

$JHK_s$ colors are a good indicator of effective temperature \citep{2014ApJ...788L..12W}, therefore, they are often used to distinguish sub-type carbon stars with different temperatures. \citet{2000MNRAS.314..630T} used the $J-H$ vs. $H-K_{\rm s}$ two-color diagram to distinguish C--N stars from C--H stars, and determined the approximate location of dC stars. \citet{2012Ap.....55..424G} showed the distribution of C--H stars and C--N stars on the $J-H$ vs. $H-K$ two-color diagram. \citet{2015RAA....15.1671S} also analyzed the distribution positions of different types of carbon stars on the $J-H$ vs. $H-K_{\rm s}$ two-color diagram, and provided the dividing line between C--H stars and C--N stars (i.e., Equation(\ref{eqjhk})).

\begin{figure}[!ht]
\plotone{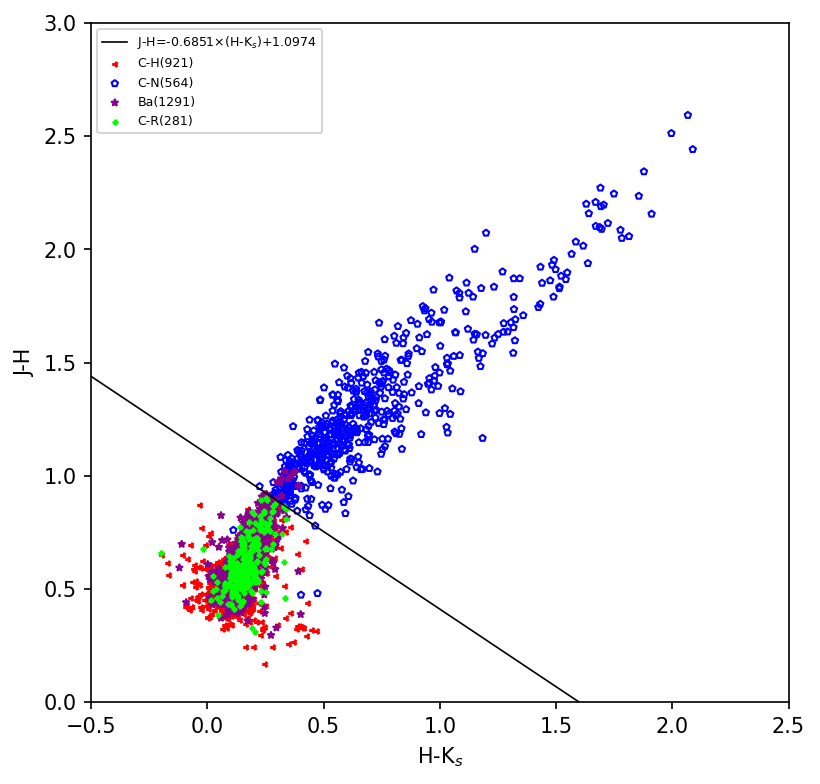}
\caption{Distribution of our carbon stars in the $J-H$ vs. $H-K_{\rm s}$ diagram. 
The black solid line (see Equation (\ref{eqjhk})) is the dividing line between the warm carbon stars (C--H, C--R, Ba stars) and the cool carbon stars (C--N stars). The blue unfilled pentagon symbols represent the C--N stars, the red dendritic symbols toward the left represent the C--H stars, the dark magenta star symbols represent the Ba stars, and the green crosses display the C-R stars.}
\label{fig:figure9}
\end{figure}

The carbon stars classified in this work are mapped in the $J-H$ vs. $H-K{\rm _s}$ plane in Figure~\ref{fig:figure9}, and the magnitudes of $JHK_{\rm s}$ come from the 2MASS database \citep{2006AJ....131.1163S}. It can be seen that, in Figure~\ref{fig:figure9}, the black straight line (Equation (\ref{eqjhk})) can separate C--H, C--R and Ba stars from C--N stars well. After statistics, the accuracy of Equation (\ref{eqjhk}) to distinguish C--N from C--H, C--R and Ba  stars are 99.9\%, 100\%, 99.2\%, respectively. There are only three C--N stars that fall on the lower left of the black solid line. We check the flag ``Qfl'' of these stars in the 2MASS catalog, which indicates the quality of the photometry in each band. It shows that except for the $K_{\rm s}$ band of, ``J055802.69+282737.4'', all the rest bands have the best quality detections, with measurement uncertainties smaller than 0.15 mag. Thus, Equation (\ref{eqjhk}) can select the C–N stars with the completeness of 98.9\% and a contamination rate of only 4.3\%, therefore, it can be used as a relatively reliable dividing line between C--N stars and the other three sub-types.

\subsection{Identification of carbon dwarfs}
\label{sec:distribution of carbon stars in the H-R diagram} 

\begin{figure}[!ht]
\plotone{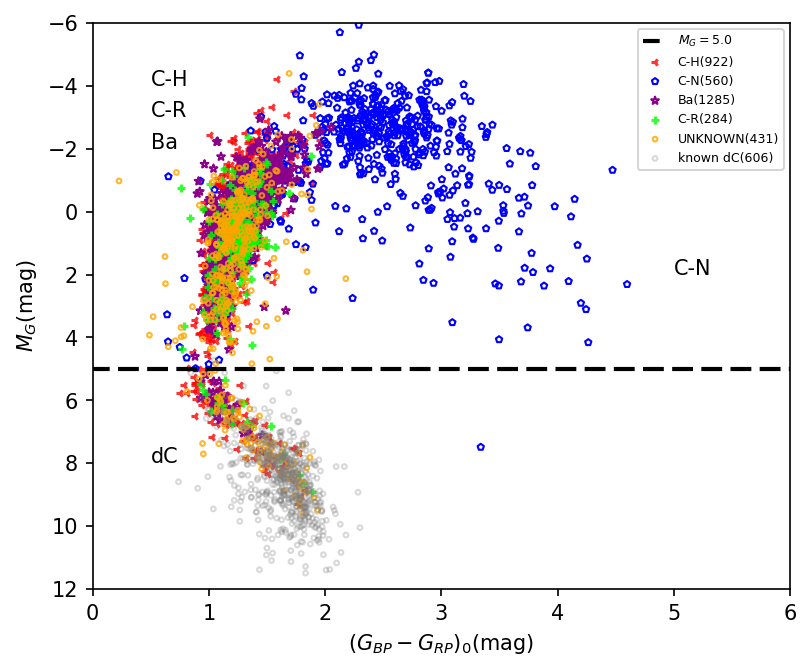}
\caption{Distribution of our carbon stars in the H-R diagram. The black 
dashed line represents the dividing line between carbon dwarfs and carbon 
giants (i.e., $M_{\rm G}$=5.0\,mag). The orange and gray circles represent ``UNKNOWN'' stars and  dC stars identified by \citet{2013ApJ...765...12G}, respectively. The other symbols are the same as in Figure~\ref{fig:figure9}. The numbers in parentheses indicate the quantities of each sub-type.}
\label{fig:figure17}
\end{figure}

Dwarf carbon (dC) stars are main-sequence stars that show
molecular absorption bands of carbon. In the mass-transfer binary (MTB), the dC progenitor accreted carbon-enhanced stellar wind material from a more massive companion that evolved into a thermally-pulsing AGB (TP-AGB) star. The dC progenitor now bears the atmospheric traces of carbon as a dC star. The TP-AGB expelled its envelope via the wind, leaving behind a white dwarf (WD), and it is now usually beyond detection in optical spectra. Previous works have shown that many carbon stars are actually carbon dwarf stars \citep{2013ApJ...765...12G,2019ApJ...881...49G,2022ApJ...926..210R}. We can easily distinguish carbon dwarfs and giants in the H-R diagram. In Figure~\ref{fig:figure17}, we show the H-R diagram constructed by $G_{\rm BP}$-$G_{\rm RP}$ and the absolute magnitude $M_{\rm G}$ of our 3482 carbon stars. We obtain the broad-band (G), the $G_{\rm BP}$-$G_{\rm RP}$ magnitude from Gaia DR3 \citep{2021A&A...649A...1G} and  distances from  \cite{2021yCat.1352....0B}. The extinction and reddening in $G$ and $G_{\rm BP}$-$G_{\rm RP}$ are corrected by adopting the $E(B-V)$ values of \cite{2019ApJ...887...93G} and extinction coefficient of \cite{2019ApJ...877..116W}. 
.

The 606 dC stars identified by \citet{2013ApJ...765...12G} are also presented in Figure~\ref{fig:figure17}, and it can be seen that the known dC stars, shown by the grey dots, are distributed in the main sequence. Then we choose a screening condition, with $M_{\rm G}$ $>$ 5.0\,mag \citep{2022ApJ...926..210R}, to select dC stars. The black dashed line in Figure~\ref{fig:figure17} represents the dividing line between carbon dwarfs and carbon giants. The lower part represents the distribution area of carbon dwarfs, while the upper part is for 
the carbon giants. We note that there is a C--N star ``J101525.92-020431.8'', in the lower area. We search for it in the SIMBAD astronomical database\footnote{http://simbad.u-strasbg.fr/simbad/}, and find that it is classified by \citet{2004A&A...418...77M} as an AGB type carbon star in the halo. Therefore, we exclude it as a dC candidate.
 
\begin{figure}[!ht]
\plotone{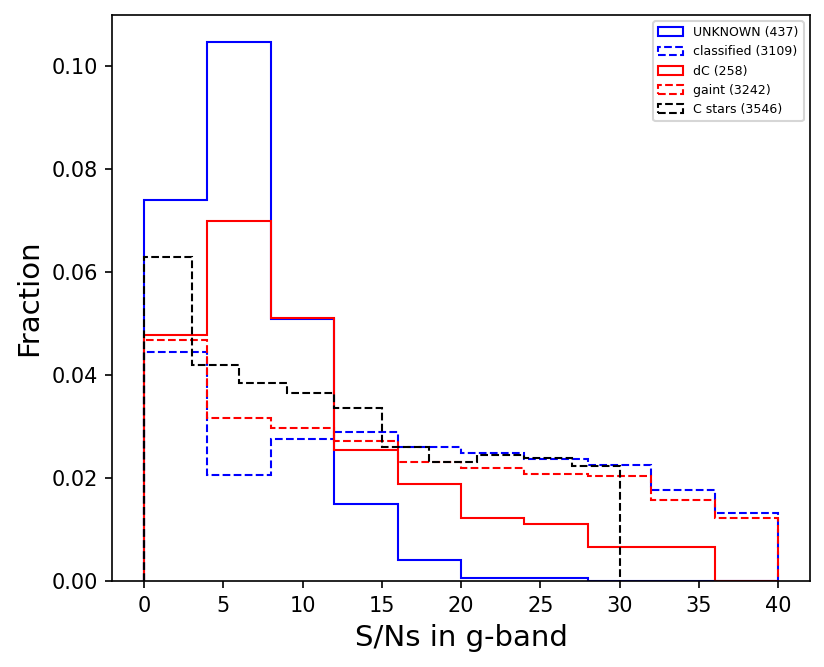}
\caption{Distribution of S/Ns in LAMOST $g$-band. The blue solid and dashed lines represent the ``UNKNOWN'' and classified carbon stars, respectively. The red solid and dashed lines represent the dCs and carbon giants, respectively. The S/N distribution of all the carbon stars is also shown by the black dashed line.}
\label{fig:snrg}
\end{figure}

Finally, we identify 258 carbon dwarf stars, including 33 Ba, 104 C--H, 21 C--R and 100 ``UNKNOWN'' stars. We note that there is a large number of dCs belong to C--H stars. This is probably because, like dCs, C--H stars are also thought to be post MTB given their orbital properties \citep{2010A&A...523A..10I}. They have suffered wind accretion from a former AGB companion, which is now a white dwarf. 
The number of dC stars classified as ``UNKNOWN'' stars is relatively larger, and the fraction of ``UNKNOWN'' for dwarfs is 40\%, it is much higher than 10\% for giants. This result might be explained by the fact that the low luminosity of dC stars leads to the low S/Ns of their observed spectra, thus, we cannot subclassify them. Since the character lines we used to classify carbon stars are mainly in the $g$-band of the LAMOST spectra, we show the distribution of S/Ns in $g$-band in Figure \ref{fig:snrg}. As expected, the S/Ns of dCs and ``UNKNOWN'' stars are statistically much lower than those of carbon giants and classified carbon stars, respectively.

\subsection{Spatial Distribution}

\begin{figure*}[!ht]
   \centering
\plotone{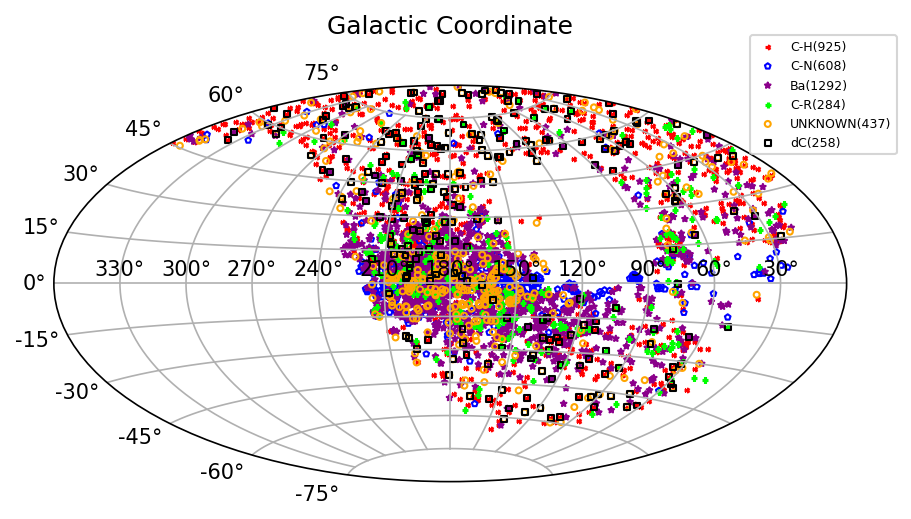}
   \caption{Spatial distribution of the 3546 carbon stars reported in this paper in the Galactic Coordinates. The black squares represent dC stars, the other symbols are the same as in Figure~\ref{fig:figure9}. The numbers in parentheses indicate the quantities of each sub-type.}.
\label{fig:figure18}
\end{figure*}

The spatial distribution in Galactic coordinates of the 3546 carbon stars is plotted in Figure~\ref{fig:figure18}. Among them, 4.6\% C--N, 63.1\% C--H, 28.2\% C--R and 16.3\% barium stars are located in the regions with $\left| b\right| \geq 30^{\circ}$. As expected, the majority of C--N (95\%)and a large number of C-R (72\%) and barium stars (84\%) are concentrated at low Galactic latitudes. A majority of the C--H stars lie in high latitudes, similar to previous works \citep{2016ApJS..226....1J,2018ApJS..234...31L}, since C--H stars are found mostly in halo populations \citep{2005MNRAS.359..531G,2010MNRAS.402.1111G}
of the Galaxy. Among the 437 ``UNKNOWN'' stars, 179 locate at $\left| b\right| \geq 30^{\circ}$, account for about 40\%. Most of the dC stars (70.9\% ) distribute at high Galactic latitudes, which supports that dC stars are from the older thick-disk and halo populations \citep{2018MNRAS.477.3801F,2022ApJ...926..210R}

\section{conclusion}
\label{sec:conclusion} 
In this work, we use the multiple line index planes, 2MASS near-infrared color-
color diagrams and manual inspections to identify carbon stars. 
We have carried out eye inspection and classification of the selected carbon 
star candidates one by one, therefore, the carbon star samples we finally obtained have the characteristics of high purity and good consistency. However, 
we lose the candidates with a small line index when cutting carbon star candidates, and they generally fall close to (0,0) in the online index plane. Compared with the identification results by the pipeline for the LAMOST DR7 spectra and those from LAMOST DR4 by \cite{2018ApJS..234...31L}, 559 and 1432 carbon stars, respectively, are unique in our catalog.

We have identified 3546 carbon stars (4542 spectra) from the low-resolution spectra of more than 10 million in LAMSOT DR7 including 925 C--H, 1392 Ba, 608 C--N and 284 C--R stars. We have labeled 437 unclassified carbon stars as ``UNKNOWN''. In the $J-H$ vs. $H-K_{\rm s}$ two-color diagram, the Equation (\ref{eqjhk}) can be used as a relatively reliable dividing line between C--N stars and other three sub-types. 

Through mapping our carbon star candidates in the H-R diagram, we have identified 258 dC star candidates from 3546 carbon stars, including 33 Ba, 104 C--H, 21 C--R, and 100 ``UNKNOWN'' stars. 

We discussed the spatial distribution of each sub-type carbon star in the Milky Way. Most of our carbon stars are distributed in the anti-galactic direction, which is caused by the LAMOST's sky survey strategy. The spatial distribution confirms that C--H stars are mostly found in halo populations. As expected, the majority of C-N, C-R, and Ba stars are distributed at low Galactic latitudes. Among the 437 ``UNKNOWN'' stars, 179 located at $\left| b\right| \geq 30^{\circ}$, account for about 40\%.

It would be very helpful for performing follow-up time domain photometric and high-resolution spectroscopic observations in the future in order to identify more carbon stars and further investigate their nature.

\begin{acknowledgments}
We thank Shuai Feng for useful discussions. study is supported by the National Natural Science Foundation of China under grants Nos. 11903012, 12173013, 11803006, 12090040, 12090044, 11833006, 12203016; the Hebei NSF No. A2019205166, A2021205006, 226Z7604G, A2022205018. We acknowledge the science research grants from the China Manned Space Project with No. CMS-CSST-2021-A09; Science and Technology Project of Hebei Education Department. The Guoshoujing Telescope (the Large Sky Area Multi-Object Fiber Spectroscopic Telescope LAMOST) is a National Major Scientific Project built by the Chinese Academy of Sciences. LAMOST is operated and managed by the National Astronomical Observatories, Chinese Academy of Sciences. 
\end{acknowledgments}


\end{document}